\documentclass[aps,prl, amsmath, showpacs, superscriptaddress, twocolumn,sort&compress,floatfix, amssymb, noeprint]{revtex4-2}
\usepackage{graphicx}
\usepackage{mathptmx, textcomp, float}
\usepackage{braket,amsfonts}
\usepackage[dvipsnames]{xcolor}
\usepackage[hidelinks]{hyperref}
\usepackage{upgreek}

\newcommand{\mm}{m_\text{eff}}

\begin{document}

\title{Floquet solitons and dynamics of periodically driven matter waves with negative effective mass}

\author{Matthew Mitchell}
\affiliation{Department of Physics and SUPA, University of Strathclyde, Glasgow G4 0NG, United Kingdom }
\author{Andrea Di Carli}
\affiliation{Department of Physics and SUPA, University of Strathclyde, Glasgow G4 0NG, United Kingdom }
\author{German Sinuco Leon}
\affiliation{Department of Chemistry, Durham University, Durham DH1 3LE, United Kingdom}
\author{Arthur La Rooij}
\affiliation{Department of Physics and SUPA, University of Strathclyde, Glasgow G4 0NG, United Kingdom }
\author{Stefan Kuhr}
\affiliation{Department of Physics and SUPA, University of Strathclyde, Glasgow G4 0NG, United Kingdom }
\author{Elmar Haller}
\affiliation{Department of Physics and SUPA, University of Strathclyde, Glasgow G4 0NG, United Kingdom }

\date{\today}

\begin{abstract}
We experimentally study the dynamics of weakly interacting Bose-Einstein condensates of cesium atoms in a 1D optical lattice with a periodic driving force. After a sudden start of the driving we observe the formation of stable wave packets at the center of the first Brillouin zone (BZ) in momentum space, and we interpret these as Floquet solitons in periodically driven systems. The wave packets become unstable when we add a trapping potential along the lattice direction leading to a redistribution of atoms within the BZ. The concept of a negative effective mass and the resulting changes to the interaction strength and effective trapping potential are used to explain the stability and the time evolution of the wave packets. We expect that similar states of matter waves exist for discrete breathers and other types of lattice solitons in periodically driven systems.
\end{abstract}

\maketitle


Ultracold quantum gases in optical lattices have proven to be excellent tools for the experimental study of novel quantum systems \cite{Lewenstein2007,Gross2017a}. In particular, optical lattices with periodic driving forces provide detailed control of tunneling between lattice sites and band structures with new, intriguing features \cite{Eckardt2017a}. Examples are the demonstration of a dynamically driven quantum phase transition between a bosonic Mott insulator and a superfluid \cite{Zenesini2009}, kinetic frustration on a triangular lattice \cite{Struck2011b}, artificial magnetic fields \cite{Aidelsburger2011,Struck2012,Parker2013,Goldman2014a}, and topological band structures \cite{Jotzu2014, Flaschner2016, Wintersperger2020a}. The systems are commonly described by the Floquet formalism, which maps the periodic driving to a time-independent Hamiltonian \cite{Holthaus2015c,Eckardt2017a}.

Essential for many experiments is the control of the dispersion relation between quasimomentum and energy in the energetically lowest lattice band \cite{Dunlap1986,Holthaus1992}. In particular, increasing the amplitude of the driving force can invert the dispersion curve, turning the energetically highest states at the edge of the 1st Brillouin zone (BZ) into the energetically lowest states \cite{Lignier2007,Eckardt2009a, Creffield2010}. The resulting evolution of atoms from the center of the BZ towards its edge \cite{Arimondo2012} is surprising because the two states in the BZ are not connected by a continuous change of driving strength. Other mechanisms such as interactions, external forces, or cooling mechanisms are required as an explanation. Understanding this time evolution is instrumental, e.g., for quantum simulation using ultracold atoms in driven systems, and the creation of Floquet condensates \cite{Arimondo2012,Heinisch2016b,Diermann2019}.

\begin{figure}[t]
\centering
  \includegraphics[width=0.49\textwidth]{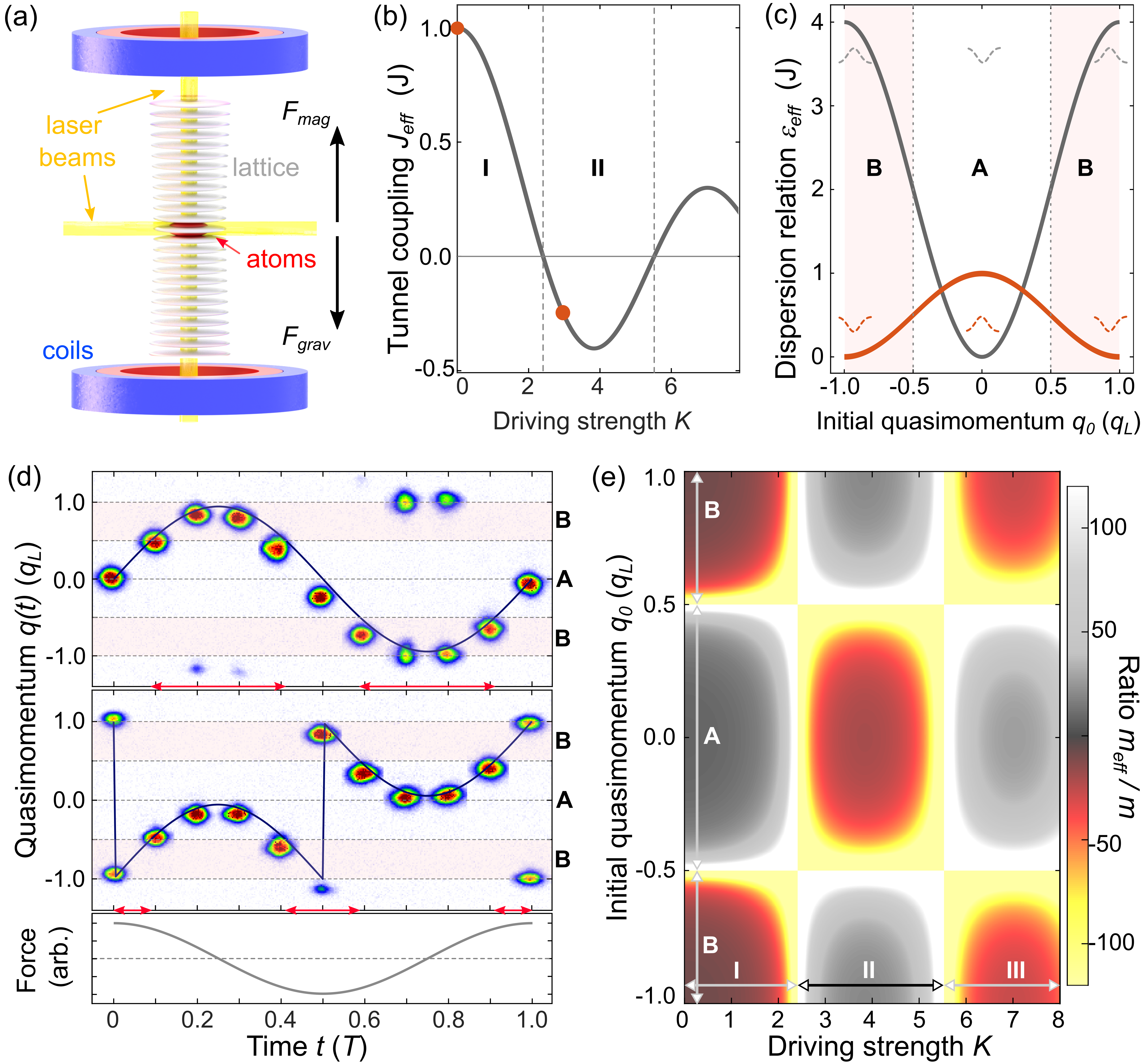}
  \vspace{-2ex}
 \caption{Experimental setup, micromotion and effective mass. (a) Experimental setup. (b) Effective tunnel coupling, $J_\text{eff}$, as a function of driving strength $K$ with circles at $K=0$ and $K=2.97$. (c) Dispersion relation for $K=0$ (gray) and $K=2.97$ (red). Dashed lines indicate the effective trapping potential in regions A and B. (d) Absorption images showing the micromotion of atoms for $q_0=0$ (top) and $q_0=q_L$ (middle) with driving force $K=2.97$ and driving period $T$. Red arrows indicate time intervals with the wave packet in region B.
 (e) Ratio of effective mass, $m_\text{eff}(K,q_0)$, to real mass, $m$ ($V=12\,E_r$). \label{fig:setup}}
\end{figure}


In this Letter, we experimentally study the evolution of a weakly interacting BEC of cesium atoms in a periodically driven 1D lattice potential after a sudden start of the driving. We demonstrate that matter waves in periodically driven systems can be stable at the center of the BZ despite an inverted dispersion curve, forming time-averaged gap solitons \cite{Hilligsoe2002a,Eiermann2004} with a negative effective mass. Those Floquet solitons have recently been demonstrated in optics using photonic wave\-guide arrays \cite{Mukherjee2020}, and they have been predicted to exist also for matter waves \cite{Konotop2002, Dabrowska2006, Kolovsky2010, Michon2018, Mukherjee2020}. We demonstrate the formation and stability of those states in momentum and position space over a duration of 1\,s.

We first provide a systematic study of the dynamics of matter waves for various driving strengths, demonstrating the stability of ground states and the decay of excited states in the BZ. We observe the typical instability of atoms at the center of the BZ for an inverted dispersion curve, however, in our weakly interacting system, this evolution towards the edge of the BZ
is caused by an external trapping potential. We provide an intuitive explanation for this effect by extending the concepts of group velocity and effective mass to periodically driven systems \cite{Arlinghaus2011d}. Atoms in states with a negative effective mass are unstable due to an effectively expulsive trapping potential along the lattice direction \cite{Pu2003b} that pushes them towards the edge of the BZ. Removing this trapping potential allows us to create and study Floquet solitons.


Our experimental starting point is a magnetically levitated BEC in a crossed-beam optical dipole trap and a vertical optical lattice potential at a wavelength $\lambda=1064$\,nm and lattice spacing $d_L=\lambda/2$ [Fig.~\ref{fig:setup}(a)]. Depending on the required driving frequency, we apply the driving force, $F(t)$, by either modulating a vertical magnetic field gradient or by periodically shaking the position of the lattice sites with detuned laser beam frequencies \cite{Arimondo2012}. For fast driving, we avoid parametric and interband excitations \cite{Weinberg2015, Reitter2017, Lellouch2017,Singh2019,Wintersperger2020b} by using shaking frequencies with excitation energies in the 1st band gap of the lattice. Further details about our setup and the parameters of our measurements are presented in the Supplemental Materials \cite{SuppMat}.


We use a semi-classical description of a wave packet in the lowest lattice band to interpret our results. A driven wave packet that is initially localized at a Bloch state with quasimomentum $q_0$ moves through the BZ according to the acceleration theorem $q(t)=q_0+\int_0^t F(t')dt'$ \cite{Arlinghaus2011d}, with $F(t)=F_0\cos(\omega t)$, where $F_0$ is the amplitude of the force and $\omega$ is the driving frequency. In the lowest band, this so-called micromotion is well-defined by the parameters $q_0$ and the dimensionless driving strength $K=F_0 d_L/(\hbar\omega)$ \cite{Arlinghaus2011d}. We show examples of this motion in Fig.\,\ref{fig:setup}(d) for $q_0=0$ and $q_0=q_L=\hbar \pi/d_L$.

Time-averaging the energy over one period of the micromotion provides an effective dispersion relation \cite{Holthaus2015c,Eckardt2017a}
\begin{align} \label{eq:dispersion}
    \epsilon_\text{eff}(K,q_0) = -2J_\text{eff}(K) \cos\left( \frac{\pi}{q_L}\,q_0 \right),
\end{align}
with an effective tunneling matrix element, $J_\text{eff}(K) = J \mathcal{J}_0(K)$, where $J$ is the tunneling matrix element and $\mathcal{J}_0(K)$ is the zeroth order Bessel function. We label the regions with positive and negative $J_\text{eff}$ with I and II ($0\le K_I<2.4 \le K_{II}\le5.5$) in Fig.\,\ref{fig:setup}(b), and the regions with quasimomenta close to the center ($|q_0|<0.5q_L$) and close to the edge ($|q_0|\ge 0.5q_L$) of the BZ with A and B in Fig.\,\ref{fig:setup}(c). The inversion of $\epsilon_\text{eff}(K,q_0)$ appears when states with $q_0$ in region A spend a large fraction of their micromotion in region B, gaining a larger average energy than states with $q_0$ in region B [e.g. top of Fig.\,\ref{fig:setup}(d)]. The argument of $\epsilon_\text{eff}$ is the initial quasimomentum $q_0$ at the start of a driving period $T=2\pi/\omega$, and experimental measurements need to probe the system stroboscopically at integer multiples of $T$.

In analogy to non-driven lattice systems, we use the effective dispersion relation to define the effective inertial mass, $m_\text{eff}$, of a wave packet
\begin{align}\label{eq:effmass}
  m_\text{eff}(K,q_0)  &=  \left[\partial_{q_0}^2 \epsilon_\text{eff}(K,q_0) \right]^{-1} \nonumber \\
                       & = \left[ 2 J_\text{eff}(K) \left(\frac{d_L}{\hbar}\right)^2 \cos\left(\frac{\pi}{q_L} q_0\right)\right]^{-1} .
\end{align}
The effective mass determines the spreading of the wave packet \cite{Morsch2006,Khamehchi2017} and its response to an external force \cite{Pu2003b}. Periodic driving always increases $|\mm|$ and it inverts the signs of $\mm$ in region II [Fig.\,\ref{fig:setup}(e)].

\begin{figure}[t]
\centering
  \includegraphics[width=0.49\textwidth]{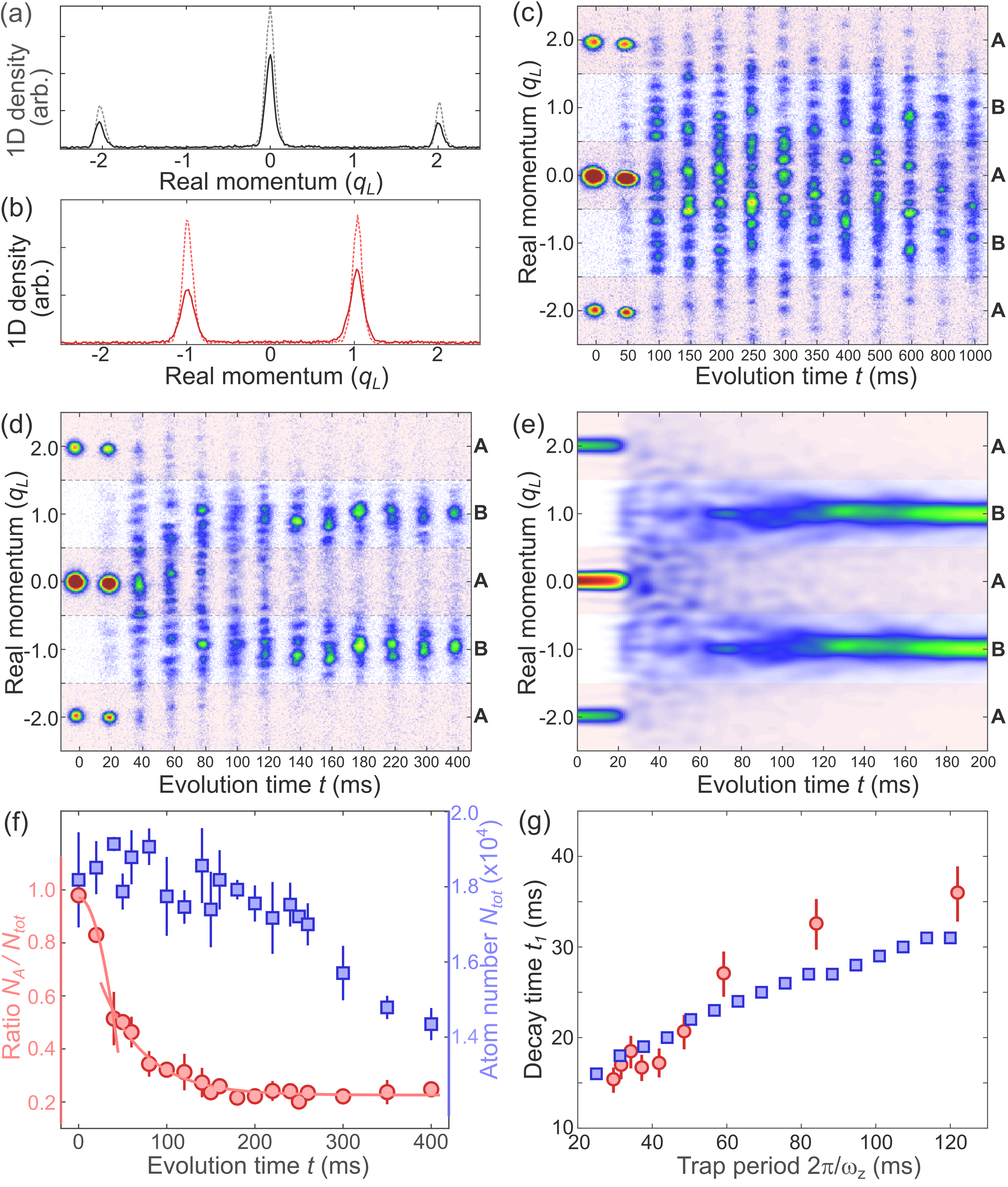}
  \vspace{-2ex}
 \caption{Time evolution of the wave packet. (a) Momentum distributions of ground states for $t=0$ (dashed lines) and $t=600$\,ms (solid lines) for $K=1.96$, $q_0=0$ and (b) for $K=2.97$, $q_0=q_L$. (c) Time evolution in momentum space with $\mm<0$ for suppressed tunneling $J_\text{eff}=-0.01J$ ($K=2.43$) and (d) for $J_\text{eff}=-0.34J$ ($K=3.26$). Red patches indicated the regions A. (e) Numerical simulation with experimental parameters of (d). (f) Evolution of $N_\text{tot}$ and ratio $R$ for the measurement in (d). Solid lines are parabolic and exponential fits \cite[Sec.\,C]{SuppMat}. (g) Decay time $t_{1}$ for increasing trap periods in experiment (red circles) and simulation (blue squares). \label{fig:evolution}}
\end{figure}

We first demonstrate the stability of the ground states for $K$ in regions I and II. The momentum distribution of the ground state in region I is centered at $q_0=0$, and we use a stationary wave packet to study the ground state for $K=1.96$. In region II, the ground state is centered at $q_0=q_L$, and we accelerate the wave packet with a magnetic field gradient to the edge of the BZ before driving with $K=2.97$ \cite{SuppMat}. The driving force is applied by shaking the lattice ($V=12.0(5)\,E_r$, $\omega=2\pi\times1\,$kHz), while keeping the vertical trapping frequency $\omega_z=2\pi\times8.2(1)$\,Hz unchanged. We measure the real momentum distribution instead of the quasimomentum to avoid difficulties when band mapping in a shaking lattice \cite[Sec.\,A]{SuppMat}. No significant change of the momentum distribution is observed over 600\,ms, except for a loss of $25\%$ of the atoms, indicating low heating rates in the ground state [Figs.\,\ref{fig:evolution}(a) and \ref{fig:evolution}(b)].


The time evolution of the state with $q_0=0$ changes drastically when we increase $K$ to cross into region II. For $J_\text{eff}\approx0$, the strong suppression of tunneling leads to dynamic localization in position space \cite{Lignier2007,Eckardt2009a,Creffield2010}. In addition, we observe the formation and spreading of a pattern of density peaks in the momentum distribution [Fig.\,\ref{fig:evolution}(c)]. Without tunneling, the system consists of an array of independent BECs, each experiencing a different time evolution of its phase. The interference pattern that forms after releasing those BECs from the lattice creates the momentum distribution, which can show revivals and temporal Talbot effects, depending on the time evolution of the phases \cite{Kaplan2000}. Such interference patterns have been demonstrated for matter waves without driving \cite{Gustavsson2011,Mark2011d}, and we believe that the density peaks in our measurement are formed by a similar mechanism in a driven system. The patterns are fluctuating for experimental runs with identical control parameters, likely from increased technical noise due to the driving force, hence we show averaged images in Fig.\,\ref{fig:evolution} (see \cite[Sec.\,C]{SuppMat} for images without averaging).

For stronger tunneling, $J_\text{eff}=-0.34\,J$ ($K=3.26$), we observe that atoms move from region A to region B at the edge of the BZ [Fig.\,\ref{fig:evolution}(d)], indicating a change of the micromotion from $q_0=0$ towards the state $q_0=q_L$ [Fig.\,\ref{fig:setup}(d)]. While it is experimentally straightforward to enforce this change with external forces, the mechanism is more intricate for periodic driving without net momentum transfer. We exclude instabilities, relaxation and energy minimization in the band as explanations by demonstrating the stability of wave packets without driving for weak interactions \cite[Sec.\,B]{SuppMat}. For a quantitative analysis of the time evolution, we measure the ratio between the number of atoms in region A and the total atom number, $R(t)=N_A(t)/N_\text{tot}(t)$. The ratio $R(t)$ decreases from close to 1 (all atoms in region A) to values between 0.5 (uniform distribution) and 0.2 (localization in region B) [Fig.\,\ref{fig:evolution}(f)].

The concept of an effective mass provides an intuitive explanation for the observed dynamics. Quenching the driving strength from $K=0$ to region II switches $\mm$ for the $q_0=0$ state from positive to negative, with important consequences for the time evolution. The evolution of a wave packet with $\mm<0$ is identical to a wave packet with positive effective mass but with sign changes of the external potential and the interactions \cite{Pu2003b,Creffield2009}. Driving with $K$ in region II effectively inverts the trapping potential for states with $q_0$ in region A.  Atoms in those states are no longer trapped by the external potential but accelerated away from the trap center. Atoms in states in region B, however, are trapped due to their positive effective mass.


To simulate the complete dynamics, we numerically integrate the discrete nonlinear Schr\"odinger equation for renormalized lattice parameters
\begin{align}
   i \hbar \partial_t \psi = (1-i\Lambda) H_\text{eff} \psi,
   \label{eq:dampedGPE}
\end{align}
with an effective Hamiltonian $H_\text{eff}$ in the tight-binding approximation \cite[Sec.\,D]{SuppMat}, a wave function $\psi$, and a phenomenological damping coefficient $\Lambda$ \cite{Choi1998, Rancon2014}. The simulation shows two stages in the time evolution [Fig.\,\ref{fig:evolution}(e)], i.e.~the initial spreading and fragmentation of the wave packet, and the subsequent slow localization of the wave packet in region B. For weak interactions, the initial spreading is dominated by the trapping potential, which accelerates the atoms and causes a rapid reduction of $R(t)$. The duration $t_{1}$ for $R(t)$ to drop to 0.7 increases with the trap period $2\pi/\omega_z$ [Fig.\,\ref{fig:evolution}(g)]. We find good agreement between simulation and experiment when we add small atom number fluctuations to the initial state to simulate finite temperature and residual non-adiabaticity during lattice loading \cite[Sec.\,D]{SuppMat}.

The second stage in the time evolution is controlled by the damping parameter $\Lambda$, which simulates energy and atom loss. Without damping, the matter waves continue to oscillate in momentum space, while damping leads to the localization in region B \cite[Fig.\,S4]{SuppMat}. We determine the $1/e$-decay time of $R$ during the second stage with an exponential fit and find good agreement between experiment and simulation for $\Lambda=0.0225$ \cite[Fig.\,S5]{SuppMat}. We believe that the energy removal in the experiment is caused by momentum-dependent atom loss. Without periodic driving, we observe an increased loss for atoms in regions with a negative effective mass and dynamical instabilities, which can cause the local collapse of the wave packet and the loss of the atoms \cite[Fig.\,S1(c)]{SuppMat}. However, with periodic driving we observe a continuous loss, which is independent of the initial state of the atoms \cite[Fig.\,S3]{SuppMat}. This loss might mask other momentum-dependent loss features that cause the damping \cite[Sec.\,D]{SuppMat}, and further studies will be necessary.


\begin{figure}[t]
\centering
  \includegraphics[width=0.49\textwidth]{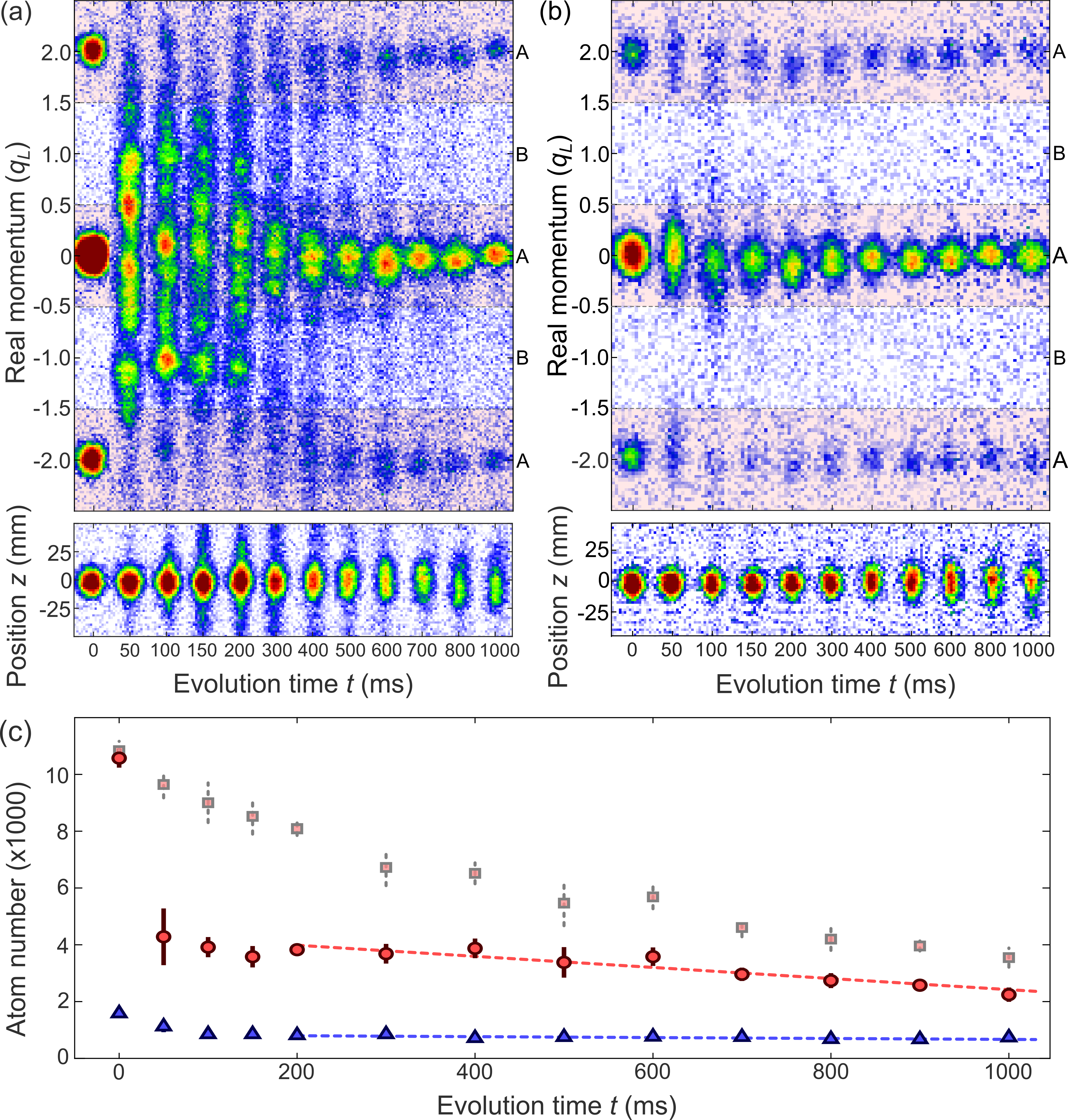}
  \vspace{-4ex}
 \caption{Matter-wave Floquet solitons. (a) Evolution of wave packet in momentum space (top) and position space (bottom) for a total initial atom number of $11,000$ and (b) of $1,600$ ($V=12.0(5)\,E_r$, $a_s=5\,a_0$, $K=4.2$). (c) Atom number in region A for $11,000$ (red circles) and 1,600 (blue triangles) initial atoms. Gray squares provide $N_\text{tot}$ for $N_\text{tot}(0)=11,000$.  All measurements are performed without vertical trapping potential. \label{fig:soliton} }
\end{figure}

Removing the trapping potential during the time evolution allows us to study the effects of weak interactions on the wave packet. A negative effective mass causes an effective sign change of the interaction strength \cite{Pu2003b,Creffield2009,Tsuji2011a}. For example, a repulsively interacting wave packet with $q_0$ in region A and $K$ in region II shows the same time evolution as a wave packet with attractive interaction and effective mass $|\mm|$. As a result, large interactions lead to dynamical instabilities \cite{Wu2001,Let2004a, Creffield2009, Lellouch2017}, while weak interactions allow us to observe stable states that show similar properties as bright matter-wave solitons with attractive interaction.

We prepare a wave packet with approximately 11,000 atoms at $a_s=5.6\,a_0$, remove the vertical trapping potential in 1\,ms, and study the dispersion of the wave packet for driving strength $K=4.2$ in momentum and position space [Fig.\,\ref{fig:soliton}(a) and (b)]. The wave packet is initially unstable and spreads over the BZ within the first 200\,ms while shedding atoms along the lattice direction. After 400\,ms, a localized wave packet forms at $q_0=0$ that is stable for the remaining observation time. The initial spread is reduced when we lower the atom number to 1,600 [Fig.\,\ref{fig:soliton}(c) and (d)], creating a wave packet that is stable both in momentum and position space for the observation time of 1\,s.

We interpret those Floquet solitons as matter-wave gap solitons that quickly cycle through the BZ. Gap solitons are bright, non-dispersive wave packets with a total energy within the band gap \cite{Hilligsoe2002a,Adhikari2009,Kizin2016}. They have been experimentally demonstrated for ultracold atoms in non-driven systems close to the edge of the band in region B where the wave packet has a negative effective mass \cite{Eiermann2004}. For periodic driving, the wave packets evolve due to the time-averaged band energy and the resulting effective mass. Controlling $\mm$ through time-averaging provides another degree of experimental control, and it allows us to create Floquet solitons in other momentum states, e.g. at $q_0=0$ for a driving strength in region II.

\begin{figure}[t]
\centering
  \includegraphics[width=0.49\textwidth]{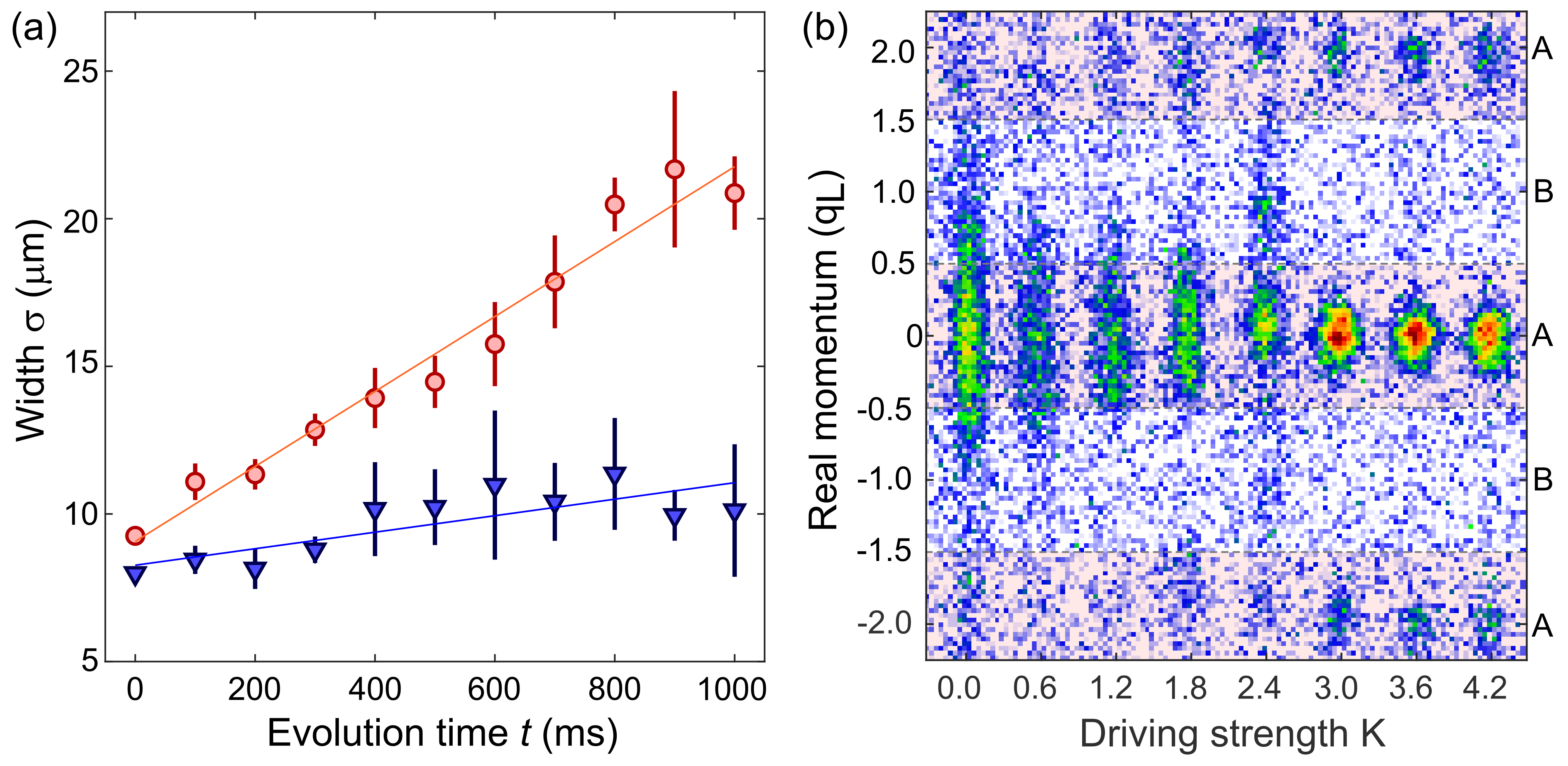}
  \vspace{-4ex}
 \caption{Matter-wave dispersion. (a) Width of the wave packet in position space in Figs.\,\ref{fig:soliton}(a) and \ref{fig:soliton}(b) for $N_\text{tot}(0) = 11,000$ (red circles) and 1,600 (blue triangles) ($V = 12.0(5)\,E_r$, $a_s = 5.6\,a_0$, $K = 4.2$). Lines are linear fits with gradients of $12.7(6)\,\upmu$m/s (red line) and $2.8(7)\,\upmu$m/s (blue line) (b) Momentum distribution after $500\,$ms of driving with for variable driving strength, $N_\text{tot}(0)=2,000$ ($V=12.0(5)\,E_r$, $a_s=3\,a_0$).
 \label{fig:soliton2}}
\end{figure}

The density profile $n(z)$ of matter-wave Floquet solitons is given by $n(z) = n_0\,\text{sech}^2 \left( \frac{z}{\sigma} \right)$, where $n_0$ is the peak 1D density and $\sigma$ is the renormalized width of the soliton that depends on the $J_\text{eff}$ and on the interaction strength \cite{Michon2018}. Using $n(z)$ as a fit function, we demonstrate that the wave packet with 1,600 atoms is almost dispersionless with a width of $\sigma=10(3)\,\upmu$m at $t=1\,$s and a dispersion of $2.8(7)\,\upmu$m/s [Fig.\,\ref{fig:soliton2}(a)]. The observed solitons are larger than expected \cite[Sec.\,E]{SuppMat}, which might be due to our limited imaging resolution, a small thermal background, or residual excitations of the soliton \cite{DiCarli2019b}.

Finally, we demonstrate the dependence of the wave packet stability on the driving strength $K$ by measuring the spread of the wave packet in momentum space after a driving duration of 500\,ms [Fig.\,\ref{fig:soliton2}(b)]. The wave packet disperses in region I ($K<2.4$) and remains stable in region II, which indicates that it is indeed the change of signs of effective mass and effective interactions that provide stability.

In summary, we study the dynamics of wave packets with negative effective mass and observe three characteristic patterns in the time evolution. First, for a strong suppression of tunneling with $J_\text{eff}\approx0$, the momentum profile shows a multitude of transient patterns which result from the phase evolution of decoupled wave packets. Second, for sufficiently strong tunneling, the wave packets accumulate at the edge of the BZ on a time scale that is set by the trapping potential along the lattice direction. We explain this effect by an inversion of the trapping potential due to the negative effective mass and by energy removal due to atom loss. Third, removing the trapping potential allows us to create Floquet solitons that are localized in position and momentum space. Their properties arise from time-averaging over the fast periodic micromotion in the first BZ, resulting in an increased stability and experimental control with  new opportunities for metrology and matter-wave quantum optics \cite{Sakaguchi2019}. We expect that other types of lattice solitons \cite{Kartashov2011a}, e.g., such as discrete breathers, have similar Floquet counterparts in periodically driven systems.


We acknowledge support by the EPSRC through a New Investigator Grant (EP/T027789/1), the Programme Grant DesOEQ (EP/P009565/1), the Quantum Technology Hub in Quantum Computing and Simulation (EP/T001062/1) and the Programme Grant QSUM (EP/P01058X/1).

\clearpage
\newpage

\renewcommand{\thefigure}{S\arabic{figure}}
\setcounter{figure}{0}

\section{Supplemental Material}

\subsection{A. Experimental setup}

Our experimental setup consists of a crossed dipole trap with vertical and horizontal laser beams at a wavelength of $\lambda=1064$\,nm [Fig.\,\ref{fig:setup}(a)]. A vertical magnetic field gradient is used to levitate the atoms against gravity, and a homogeneous magnetic field allows us to tune the s-wave scattering length $a_s$ with a magnetic Feshbach resonance \cite{DiCarli2019,*DiCarli2019c}. To study weakly interacting atoms, we create a BEC with approximately $2\times10^5$ atoms in state $F=3$, $m_F=3$ at $a_s=210\,a_0$. We reduce the atom number to $1\times10^5 - 1\times10^4$ by changing the levitating field gradient over 3\,s which tilts the vertical trapping potential and removes atoms from the trap \cite{DiCarli2019b}. Before loading the atoms into a vertical lattice (depth $V=10-12\,E_r$, $E_r=\pi^2\hbar^2/(2 m d_L^2)$, where $m$ is the mass of a cesium atom, and $d_L=\lambda/2=532\,$nm is the lattice spacing), we adjust the power in the horizontal beam to provide the desired density of the BEC with typical trap frequencies $\omega_z$ between $2\pi\times8\,$Hz and $2\pi\times34\,$Hz. The radial frequency of the vertical beam is $\omega_\rho = 2\pi\times 12(1)$\,Hz, and the s-wave scattering length $a_s$ is set to values between $3\,a_0$ and $15\,a_0$, depending on the measurement.

We create the driving force, $F(t)$, by either modulating the electrical current that generates the vertical magnetic field gradient, or by periodically shifting the position of the lattice sites \cite{Haller2010e,Arimondo2012}. The vertical lattice is formed by two counter-propagating laser beams with independently controllable frequency shifts, created by acousto-optical modulators. A frequency difference $\Delta \nu$ between the beams moves the lattice with a velocity $\Delta \nu d_L$. We create the periodic driving force within the reference frame of the lattice by modulating $\Delta \nu$ with a sinusoidal function. Alternatively, driving the atoms with a time-varying magnetic field gradient allows us to directly measure the micromotion of the atoms in the lab frame with a modulation frequency $\omega=2\pi\times50$\,Hz [Fig.\,\ref{fig:setup}(d)]. For this method, the modulation frequency is limited by the inductance of the magnetic field coils to $\omega<2\pi\times300\,$Hz. In all other measurement [Figs.\,2-4], we use the frequency shift of the laser beams to drive the system with $\omega=2\pi\times1\,$kHz. The single photon excitation energy for this frequency is within the band gap for a lattice depth $V=12\,E_r$, with band width $\approx 2\pi\times100$\,Hz and band gap $\approx2\pi\times8$\,kHz.

The atoms are detected using absorption imaging, either after a short time-of-flight of 1\,ms to allow our magnetic fields to decay [Figs.\,\ref{fig:soliton}(a) bottom, and \ref{fig:soliton}(b) bottom], or after a levitated free expansion [40\,ms in Figs.\,\ref{fig:setup}(d), \ref{fig:evolution}(c), \ref{fig:evolution}(d), 30\,ms in \ref{fig:soliton}(a) top, and 20\,ms in Figs.\,\ref{fig:soliton}(b) top, \ref{fig:soliton}(d)]. We employ band mapping with a ramp down of the lattice intensities over 2\,ms to demonstrate the micromotion in quasimomentum space for low driving frequencies in Fig.\,\ref{fig:setup}(d). However, band-mapping by ramping off the lattice intensity is ill-defined for fast driving frequencies [Figs.\,2-4] when ramp duration and driving period are comparable. Instead, we switch off the lattice beams instantly to detect the real momentum distribution of the atoms after expansion.

\begin{figure}[t]
\centering
  \includegraphics[width=0.49\textwidth]{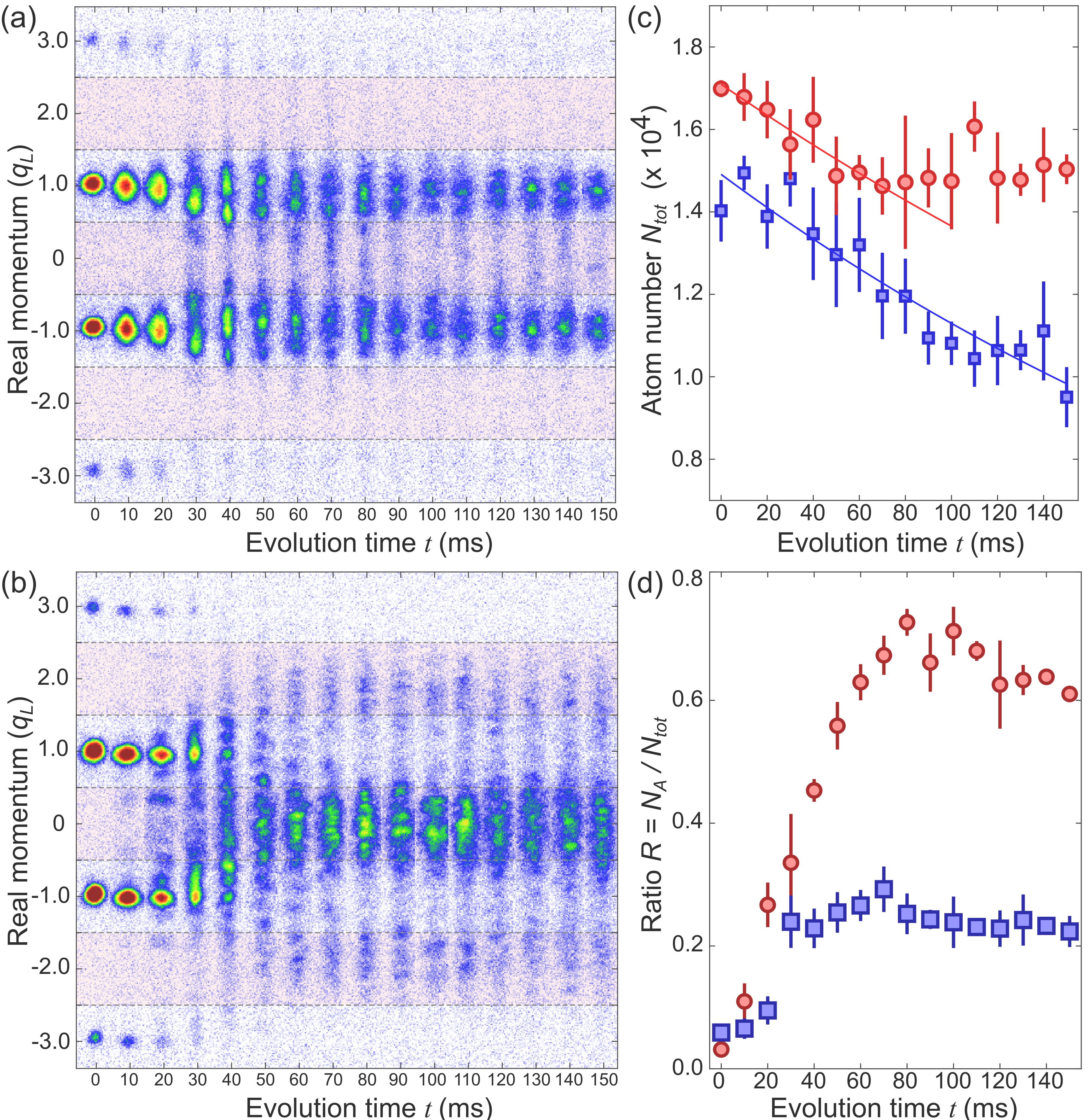}
  \vspace{-2ex}
 \caption{Time evolution of matter waves in a non-driven lattice system. (a) Evolution without the vertical confinement by the horizontal laser beam. (b) Evolution with the horizontal laser beam and $\omega_z=2\pi\times16.9(5)\,$Hz. The images are averaged over typically 3 repetitions. Red patches indicate regions with momentum $|q|<0.5$ and $1.5<|q|<2.5$. (c) Atom number for (a) blue and for (b) red. (d) Relative atom number in center of BZ for (a) blue and for (b) red. \label{fig:energy}}
\end{figure}

\subsection{B. Non-driven time evolution with $\mm<0$}

We perform an additional measurement without periodic driving to find out if a wave packet minimizes its band energy for our interaction parameters. The wave packet is initially prepared at $q=q_L$ and we study its evolution with and without an external trapping potential. Without an external trap, we do not observe an occupation of states close to the center of the BZ, which would indicate a minimization of the band energy. The distribution broadens and fragments in momentum space, but it remains in region B [Fig.\,\ref{fig:energy}(a)]. In the presence of an external trap, however, the atoms move into region A within a time interval of 70\,ms which is comparable to the trap period of $\approx60$\,ms [Fig.\,\ref{fig:energy}(b)]. We conclude that, for weakly interacting atoms, we do not observe energy minimization in the lattice band, but energy minimization in the effective trapping potential.

In detail, a BEC with approximately 16,000 atoms at $a_s=4\,a_0$ is prepared in the crossed-beam dipole trap with $\omega_{z} = 2\pi \times16.9(5)\,$Hz and the vertical lattice $V=12.0(5)\,E_r$. We ramp down the horizontal dipole beam in 3\,ms and increase the magnetic levitation field for 3\,ms to accelerate the wave packet to a quasimomentum of $q_L$. Images in Figs.\,\ref{fig:energy}(a) and \ref{fig:energy}(b) are averaged over typically 3 repetitions to account for pattern fluctuations.

The initial loss of atoms is similar for measurements with and without external trap. However, the loss slows once the atoms reach the stable region A [Fig.\,\ref{fig:energy}(c), red circles], and it continues for atoms staying in region B [Fig.\,\ref{fig:energy}(c), blue squares]. We observe that $R(t)$ increases when atoms move from region B into region A [Fig.\,\ref{fig:energy}(d)].

\begin{figure}[t]
\centering
  \includegraphics[width=0.49\textwidth]{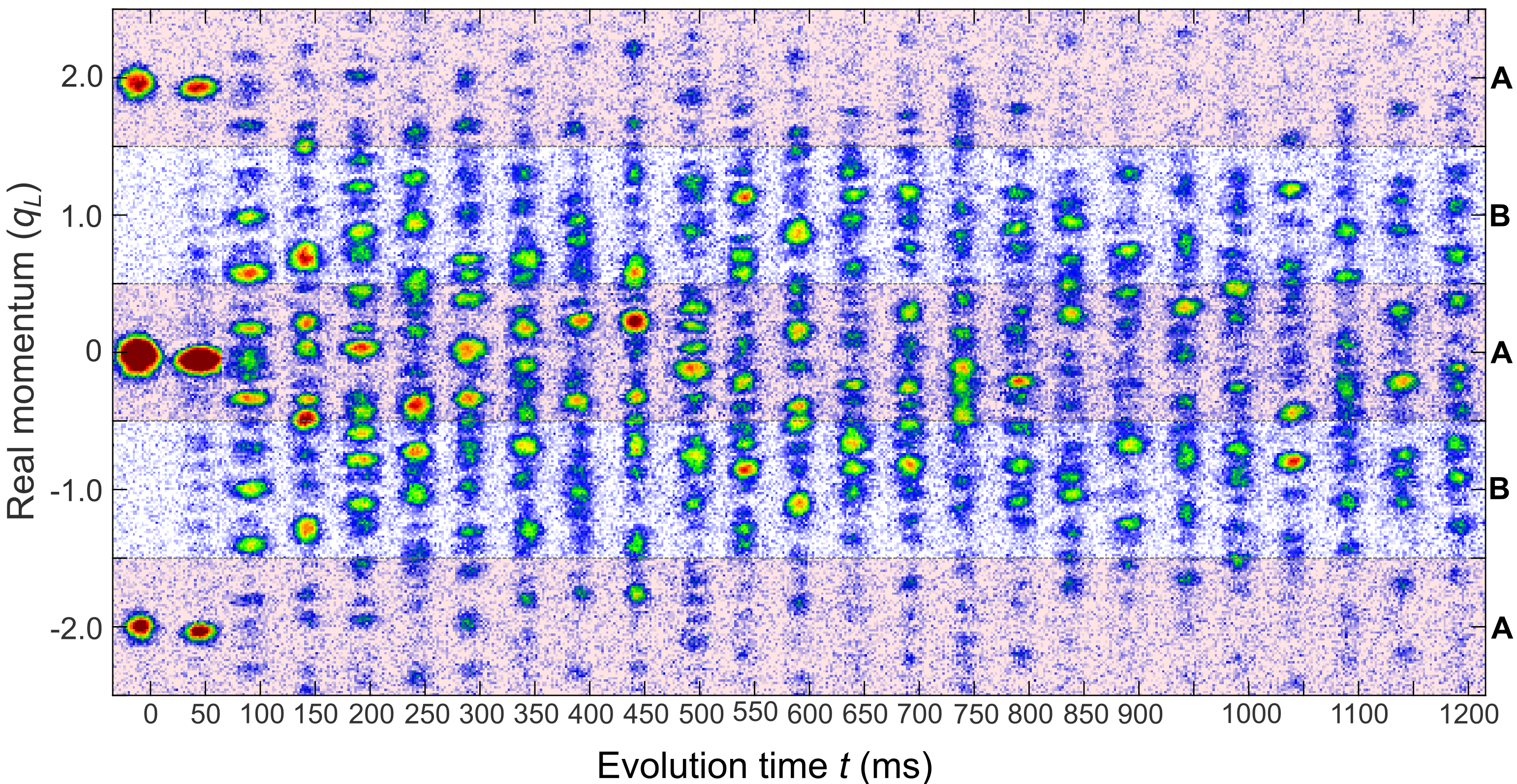}
  \vspace{-2ex}
 \caption{Time evolution of a wave packet for $\mm<0$, $K=2.43$, with $20,000$ atoms, $V=12.0(5)\,E_r$ and $a_s = 4\,a_0$ identical to the parameters used in Fig.\,\ref{fig:evolution}(c). Here, we use a single image per time step, without averaging. \label{fig:SingleShots}}
\end{figure}

\subsection{C. Driven time evolution with $\mm<0$ }

For the measurements presented in Fig.\,\ref{fig:evolution}, we prepare a BEC with approximately $20,000-30,000$ atoms in the crossed-beam dipole trap. The lattice depth is $V=12.0(5)\,E_r$ and the modulation frequency is $\omega=2\pi\times 1\,$kHz. The distribution in momentum space shows patterns of density peaks which spread within the first 150\,ms over the BZ. The patterns are fluctuating for experimental runs with identical control parameters and we show averaged images in Fig.\,\ref{fig:evolution} to indicate the variations. Images without averaging are provided for reference in Fig.\,\ref{fig:SingleShots}.

We vary the longitudinal trapping frequency $\omega_{z} = 2\pi \times 8-34\,$Hz [Fig.\,\ref{fig:evolution}(g)] and detect the atom distribution in real momentum space after $40$\,ms of levitated expansion and 1\,ms of free fall. For the scan of the trap frequency, we use a driving strength of $K=2.97$, $a_s = 7\,a_0$ and $\omega_\rho=2\pi\times 16(1)$\,Hz. For measurements in Figs.\,\ref{fig:evolution}(c) and \ref{fig:evolution}(d), we use driving strengths of $K=2.43$ and $K=3.26$, respectively, and $a_s = 4\,a_0$, $\omega_\rho=2\pi\times 16(1)$\,Hz, $\omega_z=2\pi\times 11.9(6)$\,Hz.

To model the time evolution, we define the atom number ratio $R(t)= N_A(t)/N_\text{tot}(t)$, where $N_A(t)$ is the atom number in region A and $N_\text{tot}$ is the total atom number. The decay of $R(t)$ shows two stages: a fast initial drop and a slow exponential decay. For the initial drop, we use the function $R(t) = A - B t^2$, with free parameters $A,B$, to interpolate the experimental data points, and we determine the duration $t_{1}$ with $R(t_{1}) = 0.7$ [Figs.\,\ref{fig:evolution} and \ref{fig:SuppMatt_GPEFigure_2}(a)]. These values of $t_{1}$ are shown in Fig.\,\ref{fig:evolution}(g), with the error bars indicating the duration for $R$ to drop to 0.65 and 0.75. For the slow exponential decay during the second stage, we use the fit function $A\exp(-B t)+C$ with fit parameters $A,B,C$. The $1/e$-decay times $t_2$ in Fig.\,\ref{fig:SuppMatt_GPEFigure_2}(b) are given by $1/B$.

\begin{figure}[t]
\centering
  \includegraphics[width=0.49\textwidth]{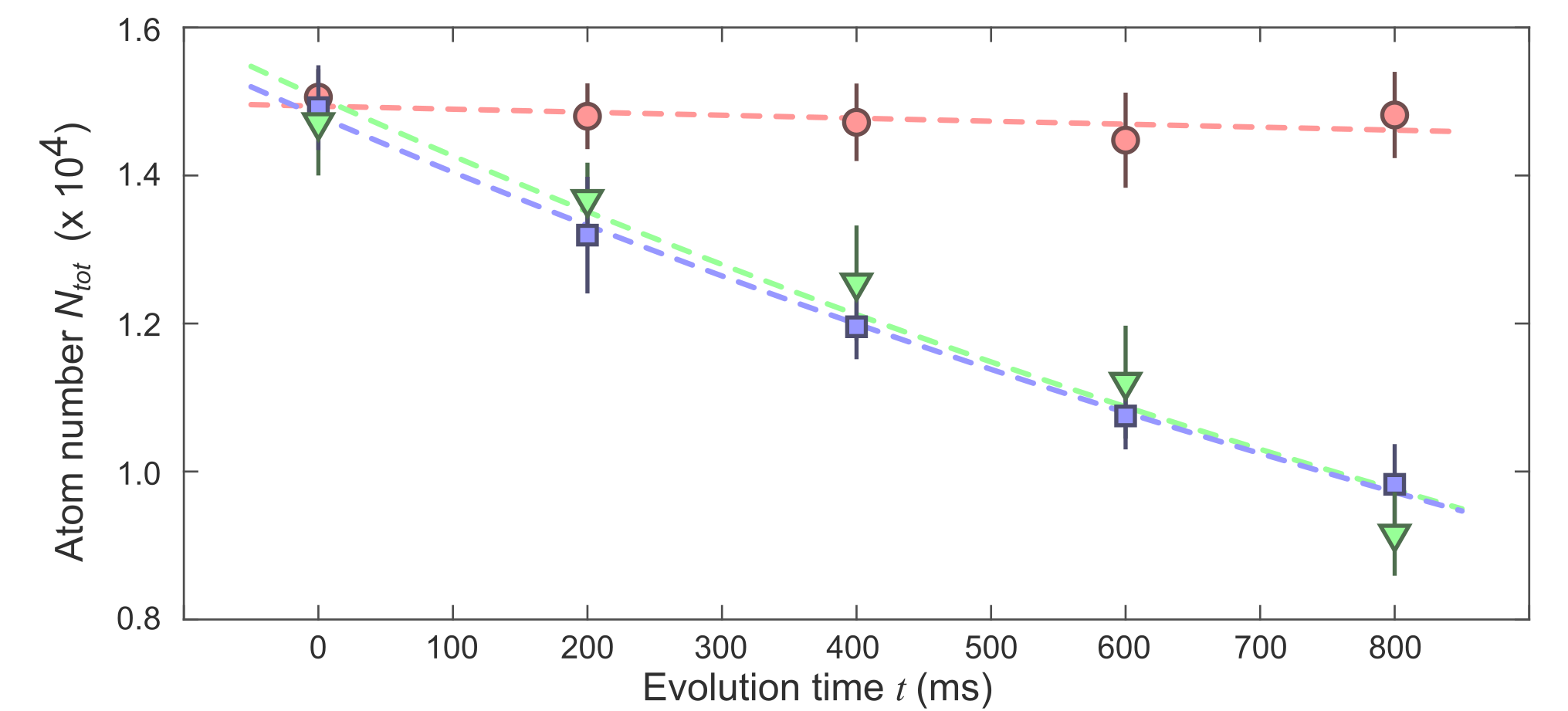}
  \vspace{-2ex}
 \caption{Atom loss with evolution time for $N_\text{tot}(0)\approx 15,000$ atoms, $V=12\,E_r$ and $a_s = 8\,a_0$. Red circles: no driving, $K=0$, $q_0=0$. Blue squares: $K=3.26$, $q_0=0$. Green triangles: $K=3.26$, $q_0=q_L$. \label{fig:Atomloss}}
\end{figure}

We determine the atom number loss during the time evolution for wave packets initially prepared at $q_0=0$ and $q_0=q_L$. Wave packets with $q_0=0$ have an excess energy $4 J_\text{eff}\approx 2\pi\hbar \times 20\,$Hz compared to final states with $q_0=q_L$, and we believe that the energy is removed through loss of atoms during the time evolution. The measured exponential lifetimes are 1.9(1)\,s ($q_0=0$) and 1.8(3)\,s, ($q_0=q_L$), and $>30$\,s without a driving force ($K=3.26$, $V=12.0(5)\,E_r$, $\omega_z= 2\pi\times 8.2(1)\,$Hz, and $a_s = 8\,a_0$) ) [Fig.\,\ref{fig:Atomloss}]. We observe an increase of atom loss due to the driving force, but no significant difference between wave packets that evolve from the center of the BZ to the edge, and wave packets that already start at the edge. We conclude that energy gain and atom loss due to the inverted dispersion relation is negligible compared to heating and atom loss due to the periodic driving force.

\subsection{D. Damped non-linear Schr\"odinger equation}

We use a damped 1D Gross-Pitaevskii equation to model the time evolution of the condensate following the quench of the shaking \cite{Choi1998}:
\begin{equation}
    i \hbar \partial_t \psi = (1-i\Lambda) H_{\textrm{eff}} \psi, \label{eq:dissipativeGPE}
\end{equation}
with the complex vector $\psi$ whose elements are the amplitudes of the condensate wave function at each lattice site. The non-linear effective Hamiltonian, $H_{\textrm{eff}}$, and details of $\psi$ are explained below. The parameter $\Lambda$ in Eq.~(\ref{eq:dissipativeGPE}) introduces a dissipation process that, heuristically, explains the loss of energy after quenching the driving. The dissipative model Eq.~(\ref{eq:dissipativeGPE}) has been employed to study non-equilibrium processes in BECs, such as the stability of topological excitations \cite{Yakimenko2013} and the dynamics following a quench \cite{Rancon2014}. In our case, Eq.~(\ref{eq:dissipativeGPE}) with a constant and spatial-independent $\Lambda$ explains qualitatively our observations.

We apply the tight-binding approximation to derive the 1D effective Hamiltonian, $H_{\textrm{eff}}$ \cite{Smerzi2003}. With this approach, the condensate wave function is modelled as a superpositions of Gaussians centred at each node, $j$, of the optical lattice:
\begin{equation}
\psi(\vec{r},t) = \left(\frac{m \bar{\omega}}{\hbar} \right)^{3/4}\sum_{j} \psi_j(t) \exp\left[ -\frac{(z-j d_L)^2}{\tilde{a}_{z}^2}\right] \exp\left[ -\frac{\rho^2}{a_{\rho}^2} \right],
\label{eq:wavefunction}
\end{equation}
where where $\psi_j$ is the complex amplitude of the wave function at lattice site $j$, $d_L$ is the periodicity of the lattice, $a_{\rho}=\sqrt{\hbar/m \omega_{\rho}}$ and $\tilde{a}_{z}=\sqrt{\hbar/m \tilde{\omega}_z}$ are the radial and axial Gaussian widths with trap frequencies $\omega_{\rho}$ and $\tilde{\omega}_{z}$ at the center of the nodes of the optical lattice, respectively. $\bar{\omega} = (\tilde{\omega}_z\omega_{\rho}^2)^{1/3}$ is the geometrical average of the trapping frequencies at each lattice site. The wave function Eq.\,(\ref{eq:wavefunction}) is normalized, $\sum_j \left| \psi_j\right|^2 = 1$.

\begin{figure}[t]
\centering
  \includegraphics[width=0.49\textwidth]{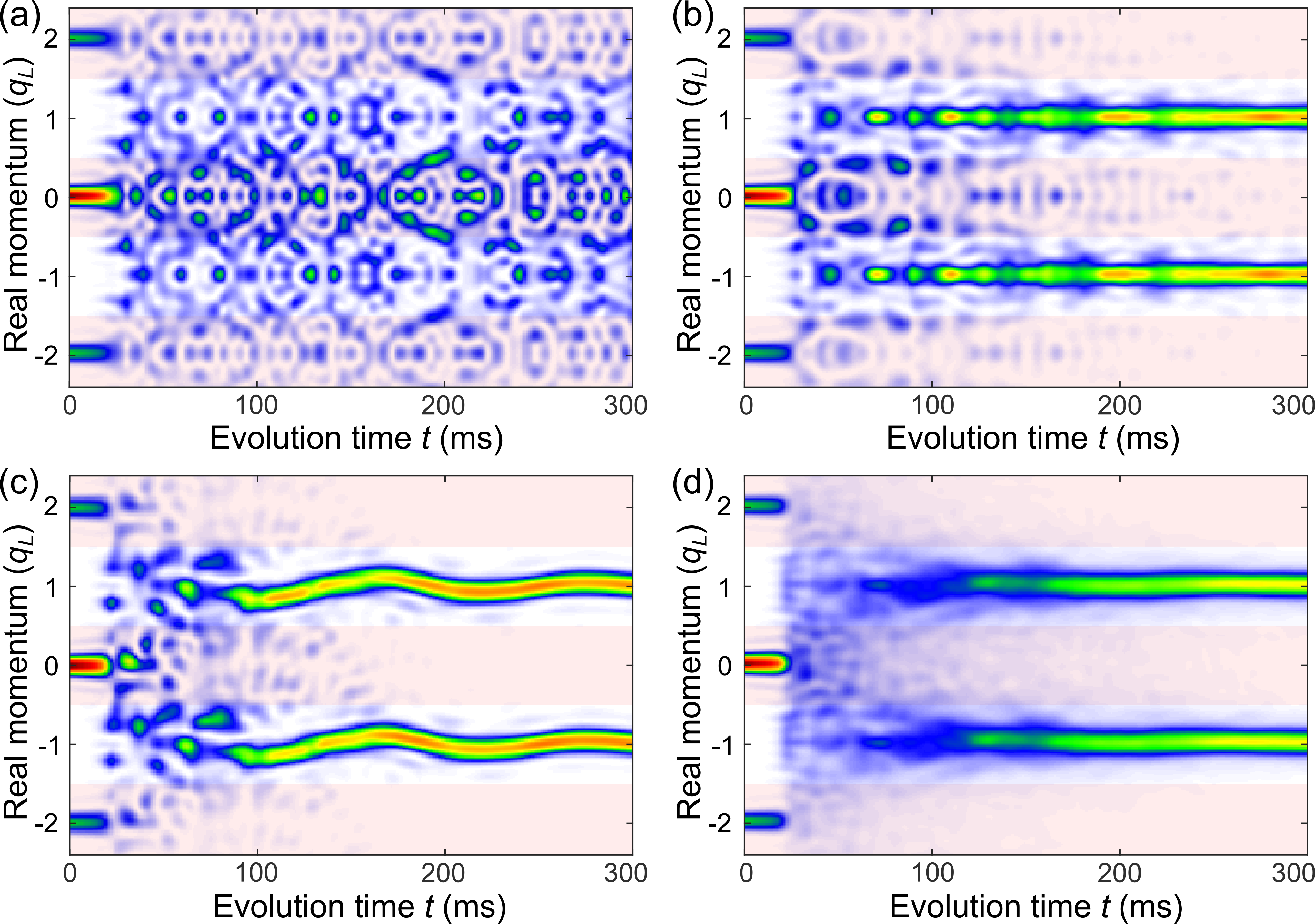}
\caption{\label{fig:SuppMatt_GPEFigure}  Simulation of the time evolution of a wave packet using Eq.~(\ref{eq:dissipativeGPE}) with $N=24,000$, $a_s=4.0\,a_0$, $\omega_z=2\pi\times29.2~$Hz, $\omega_\rho= 2\pi\times16~$Hz, $V=12\,E_r$. The driving frequency is $\omega = 2\pi\times 1~$kHz and the driving strength is $K=3.26$. (a) No damping, $\Lambda = 0$. (b) With damping, $\Lambda = 0.0225$. (c) Single realization and (d) an average over 32 realizations with damping and atom number fluctuations.}
\end{figure}

Including the harmonic trapping potential
\begin{align*}
    V(\rho,z) =  \frac{1}{2} m (\omega_{\rho}^2 \rho^2 + \omega_z^2 z^2)
\end{align*}
and the mean field interaction $\frac{4 \pi \hbar^2 a_s N}{m} \left| \psi(\vec{r},t)\right|^2$, the factorization Eq.~(\ref{eq:wavefunction}) leads to the mean-field Hamiltonian \cite{Smerzi2003}:
\begin{eqnarray}
H_{\textrm{eff}} &=& \sum_j \left\{ \epsilon_j \psi_j^* \psi_j  -   \mathcal{J}_0(K)  \left( J_j^+\psi_j^* \psi_{j+1} + J_j^- \psi_j^* \psi_{j-1} \right) \right. \nonumber \\
& & \left. + g |\psi_j|^2 -  \chi \left[ |\psi_j|^2 \psi_j ( \psi_{j+1}^* + \psi_{j-1}^*) + \textrm{c.c.} \right] \right\}, \nonumber \\
\end{eqnarray}
where $\mathcal{J}_0(K)$ is the zero-order Bessel function of the first kind and:
\begin{eqnarray}
\epsilon_j &=& \frac{\hbar \omega_z}{2} \frac{\omega_z}{\tilde{\omega}_z}  \left(\frac{d^2}{\tilde{a}_z^2}j^2 + \frac{1}{2}\right) + \frac{\hbar \omega_\rho \sqrt{\pi}}{2}\nonumber \\
J_j^{\pm} &=&  J + \left[\frac{m \omega_z^2 d_L^2 \sqrt{\pi}}{2} \left(j^2\mp j+\frac{\tilde{a}_z^2}{2 d_L^2}\right)  \right. +  \nonumber \\
& & \left. \frac{\hbar \omega_\rho \pi}{2} \right] \exp\left(- \frac{d_L^2}{4\tilde{a}_z^2}\right) \nonumber \nonumber \\
g    & = & \frac{}{}\frac{1}{(2\pi)^{3/2}}\frac{\tilde{a}_{z}}{a_{\rho}^2} \frac{4 \pi \hbar^2 a_s N}{m} \nonumber \\
\chi & = & \frac{4 \pi \hbar^2 a_s N}{m} \left( \frac{\pi}{2}\right)^{3/2} \left(\frac{m \bar{\omega}}{\hbar}\right)^3   \exp \left( -\frac{7d_L^2}{8 \tilde{a}_z^2}\right). 
\nonumber
\end{eqnarray}
The spatial dependence of the energy shift, $\epsilon_j$, and the tunnelling rate, $J_j^{\pm}$, reflect the lattice-symmetry breaking effects of the trap.

In the non-dissipative case ($\Lambda =0$), we observe that the initial state, centered around $q_0=0$, evolves towards the edge of the BZ as in the experiment. In this case, the central region and the edges of the BZ become equally populated and the atoms never localize in region B [Fig.\,\ref{fig:SuppMatt_GPEFigure}(a)]. Setting $\Lambda > 0$, we observe a significant change of the dynamics [Fig.\,\ref{fig:SuppMatt_GPEFigure}(b)].  The recurrences of the wave function disappear (due to the loss of energy) which leads to, after a transition period, the settling of the wave function at the edge of the BZ ($q_0=q_L$). For our experimental parameters, a value of $\Lambda=0.0225$ provides a time evolution that is similar to our experimental measurements, except for a time delay of the initial spreading of the wave packet. We reduce this initial time delay by adding small atom-number fluctuations to the initial state \cite{Michon2018}. The fluctuations are implemented by shifting the ground state wave function at each lattice site by a random real number from a uniform distribution centred at zero and with amplitude $\Delta \psi = 0.0005$. Results of a typical single run and after taking the average over several realisations with randomized noise are shown in Figs.\,\ref{fig:SuppMatt_GPEFigure}(c) and \ref{fig:SuppMatt_GPEFigure}(d), respectively.

\begin{figure}[t]
\centering
  \includegraphics[width=0.46\textwidth]{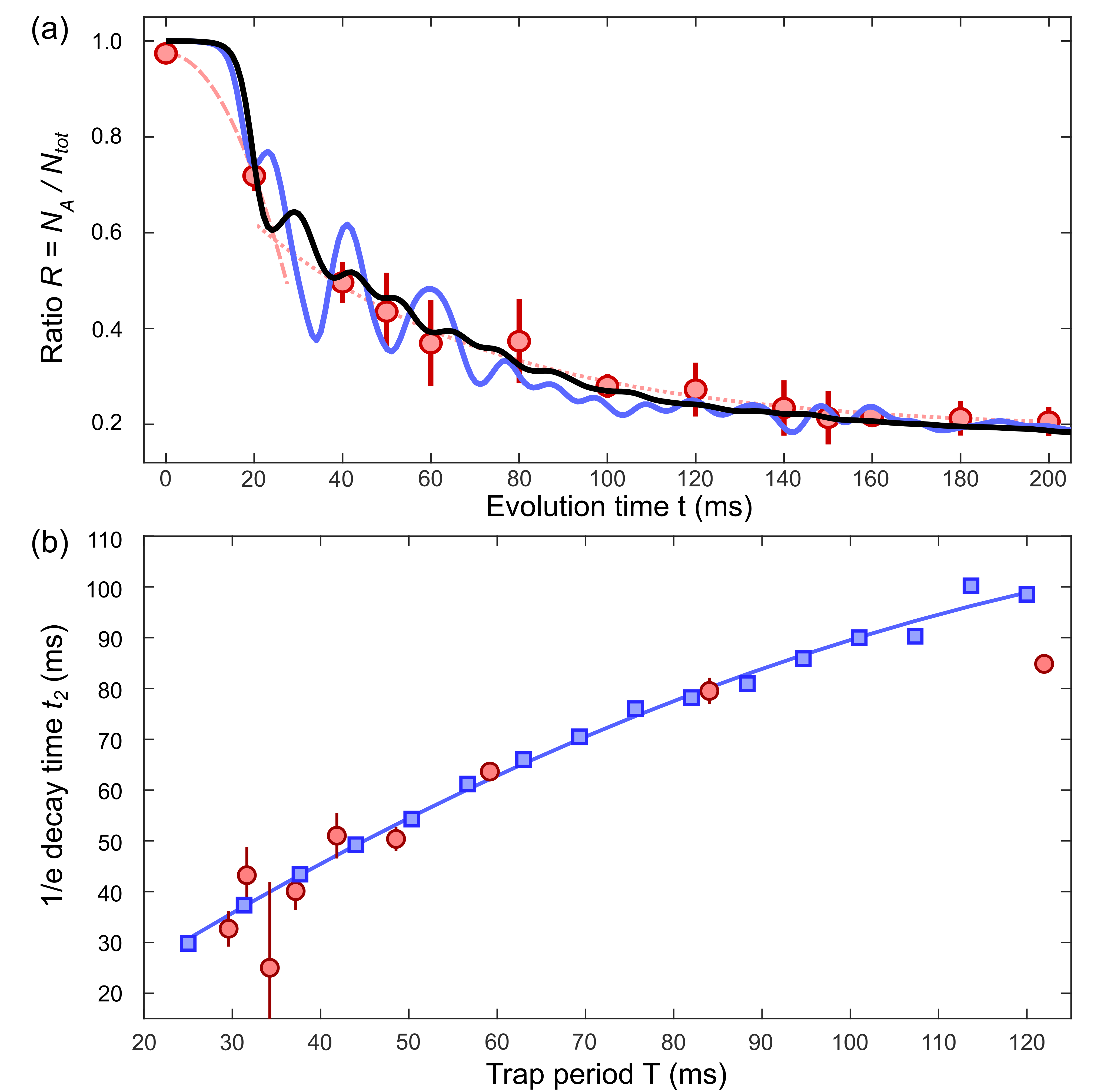}
\caption{\label{fig:SuppMatt_GPEFigure_2} (a) Comparison of the decay of the number of atoms in region A, $R=N_A/N_{\textrm{tot}}$, as a function of time. The red circles are experimental measurements with $a_s=7.0\,a_0$, $\omega_z = 2\pi \times 20.6(6)~$Hz, $\omega_{\rho} = 2\pi\times 16(1)~$Hz, $V_0 = 12.0(5)~E_r$ and $K=2.97$. The red dotted (dashed) line fits the late (early) decay of $R(t)$ to an exponential (quadratic) function, defining a slow (fast) times scale [Sec.~D and E]. The blue and black lines are numerical results of a single run and of an average over 32 realizations of the atom number fluctuations, respectively. (b) Comparison of the experimental (red circles) and numerical (blue squares) $1/e$-decay times during the second stage of the time evolution for varying trap period. The numerical results are obtained using the experimental parameters and $\Lambda = 0.0225$, $\Delta \psi = 0.0005$ and $N=24,000$. The line is a quadratic fit to the simulation data to guide the eye.}
\end{figure}

Our numerical results show that the trajectory of the fraction $R(t)$ of the cloud in region A has two time scales, as shown in Fig.\,\ref{fig:SuppMatt_GPEFigure_2}(a) for a representative set of parameters. Initially, the transition from $R=1$ to $R\approx 0.7$ occurs quickly defining a fast time scale that is dominated by the trap period. We compare the duration $t_{1}$ with $R(t_{1}) = 0.7$ in simulation and experiment and find good agreement [Fig.\,\ref{fig:evolution}(g)].

The second stage of the time evolution follows an exponential decay that depends on $\Lambda$ and on the trap period. The values of $\Lambda=0.0225$ and of $\Delta \psi = 0.0005$ are chosen to match the experimental data. We find good agreement of the $1/e$-decay time $t_2$ between the simulation and the experimental data for varying trap periods [Fig.\,\ref{fig:SuppMatt_GPEFigure_2}(b)]. The damping parameter $\Lambda$ is constant in the simulation, but it might decrease in the experiment for atoms in stable states at the edge of the BZ. We speculate that deviations between experiment and simulation in Figs.\,\ref{fig:evolution}(g) and \ref{fig:SuppMatt_GPEFigure_2}(b) are caused by state-dependent damping mechanisms in the experiment.

\subsection{E. Floquet solitons with periodic driving}

For the measurements presented in Fig.\,\ref{fig:soliton}, we prepare BECs with approximately $11,000$ and $1,600$ atoms a 1D lattice with $V=12.0(5)\,E_r$, $\omega_z = 2\pi\times 23.9(4)\,$Hz, and $a_s = 5.6\,a_0$. The radial trap frequency of the vertical laser beam is $2\pi\times12(1)\,$Hz. We remove the horizontal laser beam in 1\,ms to allow for a vertical expansion, and shake the lattice for the evolution time $t$ with $\omega = 2\pi\times 1\,$kHz and $K=4.2$. Figures\,\ref{fig:soliton}(a) and \ref{fig:soliton}(b) show the momentum distribution of the evolving wave packet. The images are averaged over typically 4 repetitions.

We measure the spread of the wave packets in positions space [Fig.~\ref{fig:soliton2}(a)], using the expected density profile, $n(z)$, for a Floquet soliton \cite{Michon2018}
\begin{align*}
    n(z) = n_0\,\text{sech}^2 \left( \frac{z}{\sigma} \right)
\end{align*}
as a fit function. Here, $n_0$ is the peak 1D density and $\sigma$ is the width of the soliton. The width $\sigma$ increases slowly with $12.7(6)\,\upmu$m/s for $N_0 = 11,000$, and the wave packet is almost dispersionless with $2.8(7)\,\upmu$m/s for $N_0 = 1,600$. For $N_0 = 1,600$ and $t=1\,$s, we measure a width $\sigma=10(3)\,\upmu$m and a peak density $n_0\approx20$\, $\text{atoms}/\upmu$m. The predicted width of the gap soliton is
\begin{align*}
    \sigma = \sqrt{ \frac{-2 J_\text{eff}(K)  d_L}{U n_0} }= 4(1)\,\upmu\text{m},
\end{align*}
where $U$ is the on-site interaction strength \cite{Michon2018}. The predicted value is below our resolution limit ($\approx7\,\upmu$m for distinguishable objects) and smaller than our measured value. We believe that the deviation is caused by a broadening due to our imaging resolution, by a small thermal background, and by residual excitations of the soliton \cite{DiCarli2019b}.

\bibliography{NegMassBib}

\end{document}